\documentclass[11pt,a4paper]{article}   
\usepackage{moriond,epsfig}

\newcommand{\beq}{\begin{equation}}    
\newcommand{\eeq}{\end{equation}}    
\newcommand{\bea}{\begin{eqnarray}}    
\newcommand{\beas}{\begin{eqnarray*}}    
\newcommand{\beau}[1]{\begin{equation} \label{#1} \begin{array}{rcl}}    
\newcommand{\eea}{\end{eqnarray}}    
\newcommand{\eeas}{\end{eqnarray*}}    
\newcommand{\eeau}{\end{array} \end{equation}}    
\newcommand{\bay}{\begin{array}}    
\newcommand{\eay}{\end{array}}    
\newcommand{\ds}{\displaystyle}

\newcommand{\lora}{{\longrightarrow}}

\newcommand{\ra}{{\rightarrow}}

\newcommand{\pcut}{p_{cut}}
\newcommand{\psat}{p_{sat}}
\newcommand{\Njet}{N_{jet}}

\begin{document}   
\vspace*{2cm} 
\title{SEMI-HARD SCATTERINGS AT RHIC AND LHC: \\ 
INITIAL CONDITIONS AND CHARGED MULTIPLICITIES
\footnote{Talk given at the {XXXVI rencontres de Moriond: QCD and hadronic
interactions}, Les Arcs (FRA) 17-24 March 2001.}}    
   
\author{ A. ACCARDI }   
   
\address{Universit\`a di Trieste, dipartimento di fisica teorica, \\    
strada Costiera 11, I 34014 Trieste, ITALY \\
{\rm and} INFN sezione di Trieste\\
e-mail: accardi@ts.infn.it} 
   
\maketitle\abstracts{ 
Minijet production in ultra-relativistic heavy-ion collisions is discussed by
taking semi-hard parton rescatterings explicitly into account. 
At both RHIC and LHC energies we find sizable effects on global
characteristics of the nuclear collision like the initial multiplicity
and transverse energy of the minijets. The dependence of these
quantities on the cutoff that separates soft and hard
interactions becomes much smoother after the inclusion of the rescatterings. 
This allows to define an energy and centrality dependent {\it saturation
cutoff} and to push perturbative computations to rather low values of the
cutoff. As an application we compute the charged multiplicity at
mid rapidity and compare it to the recent RHIC data. }    
   
In nuclear collisions at ultra-relativistic energies the initial production 
mechanism is the liberation of a great number of partons from the nuclear 
wave-functions in a very short time from the beginning of the collision. 
Then, this dense system of partons, also called {\it minijet plasma},
will then  evolve, possibly thermalizing and giving rise to the quark
gluon plasma.  
The signals of its creation are very sensitive to the minijet plasma initial 
conditions, that in turn can be related to final state bulk observables 
like the charged particle multiplicity and their transverse energy. On
the other hand, at RHIC and LHC an increasingly large fraction of the  
event is expected to be due to hard and semi-hard interactions, so
that we can try to gain control on the initial conditions by using
perturbative QCD. 

\begin{figure} 
\begin{center}
a) \epsfig{figure=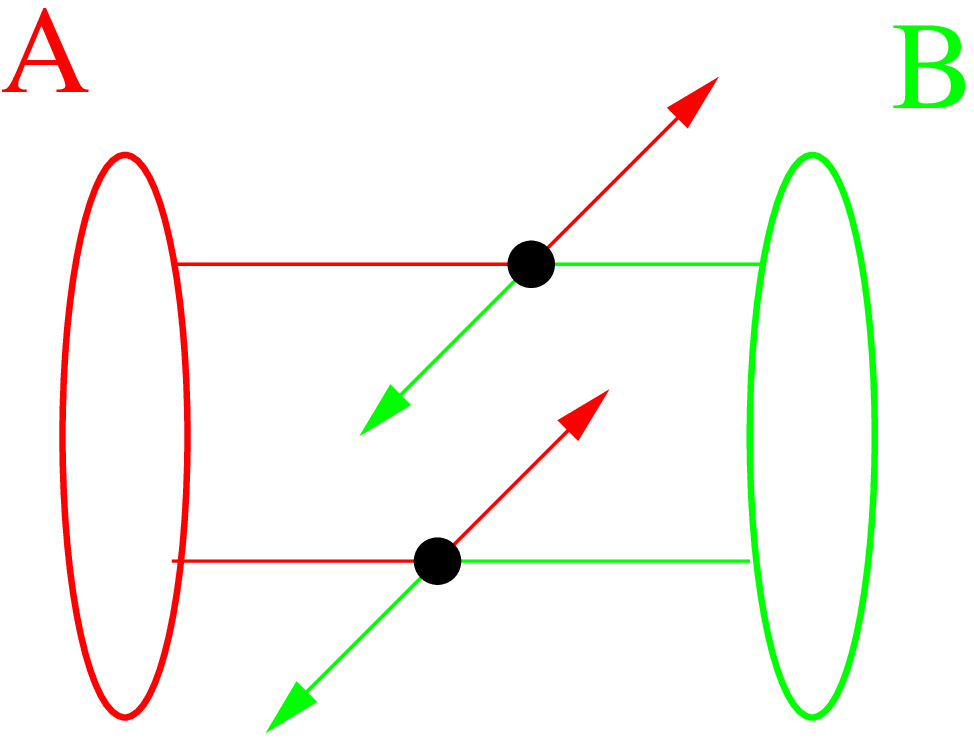,height=2.5cm} 
\hspace*{2cm}
b) \epsfig{figure=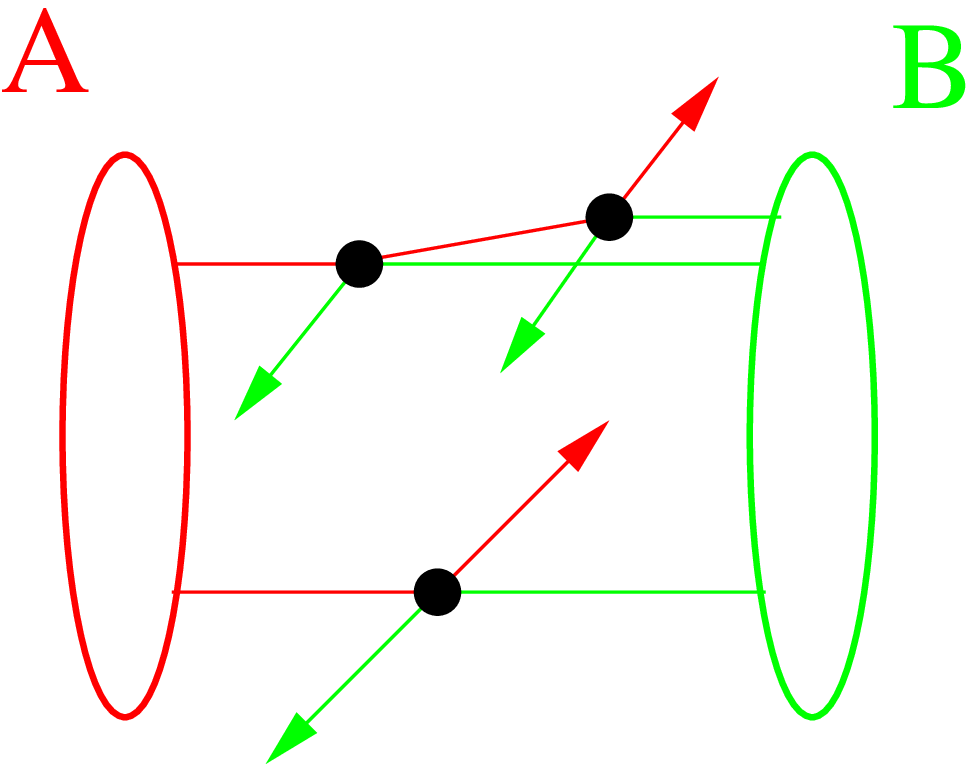,height=2.5cm} 
\vspace*{-.2cm}
\parbox{15cm}{\caption{a) An example of disconnected collisions. b)
Rescatterings.} 
\label{fig:collisions} } 
\end{center}
\vspace*{-.2cm}
\end{figure}

The usual way to do this is to take into account the Eikonalized minijet 
cross-section $ \sigma_{mj} = \int d^2\beta \, 
\left( 1 - \exp\left[- \sigma_J  T_{AB}(\beta) \right] \right) $,  
where $\beta$ is the nuclear impact parameter, 
$T_{AB}(\beta) = \int d^2b \, \tau_A(b-\beta)\tau_B(b)$, 
$\sigma_J = \int dxdx'G(x)\sigma_H(xx') G(x')$, 
$G$ is the parton distribution function and $\tau_{A(B)}$ are the
nuclear~thickness~functions; $\sigma_H(xx')$ is the perturbative 
parton-parton cross section defined with an infrared cutoff $p_{cut}$, that
separates soft and semi-hard scatterings; here and in the following the flavour
indices are suppressed for simplicity.
In this way we are describing different parton pairs interacting in different 
points in the plane transverse to the beam; we will call these events
{\it disconnected collisions}, see Fig. \ref{fig:collisions}a. In each
collision two back-to-back jets are produced, so that the minijet
multiplicity is easily seen to be 
\beq
	N_{jet}^{eik}(\beta) = 2N_{col}(\beta) = 2 \sigma_J T_{AB}(\beta)
		= 2\int dxd x' d^2b \, \Gamma_A(x,b-\beta) \sigma_H(xx') 
		\Gamma_B(x',b) \ ,
  \label{Neik}
\eeq
where $\Gamma_A(x,b) = G(x)\tau_A(b)$. 

There are two problems with this expression. First, it diverges when 
$p_{cut}\ra 0$ due to the inverse power divergence of the perturbative 
cross-section $\sigma_H$. Second, at very high energy and large atomic 
numbers a typical projectile parton is traversing a dense target
nucleus, so that the chance  
of scattering more than once may become 
non negligible. To have a quantitative feeling on this, one can take the 
average number of parton-parton collisions in (\ref{Neik}), drop the
integrations over  
$x$ and $b$, divide by the number of incoming partons, $\Gamma_A$, and obtain 
the average number of collisions per incoming parton, 
$\langle N_{scat}(x,b) \rangle = \int dx' \sigma_H(xx') \Gamma_B(x',b) $. 
For example, at LHC with a cutoff $p_{cut}=2$ GeV a parton in the central 
rapidity region suffers on average two to three collisions over most of the 
target transverse area \cite{AT}. In conclusion, parton rescatterings
must be taken  into account in the description of the collision dynamics.

\begin{figure} 
\begin{center}
a) \epsfig{figure=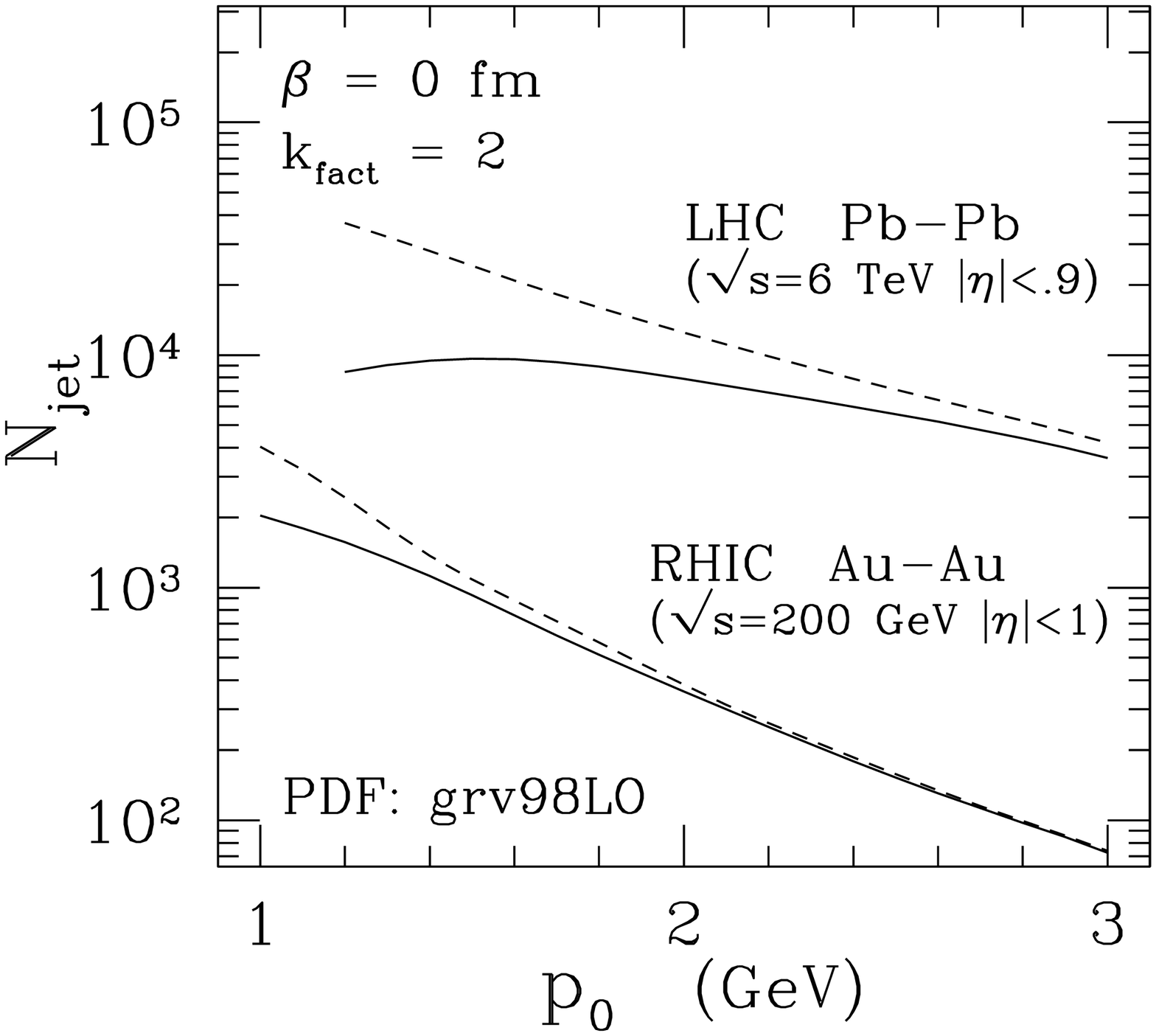,height=4.5cm}
b) \epsfig{figure=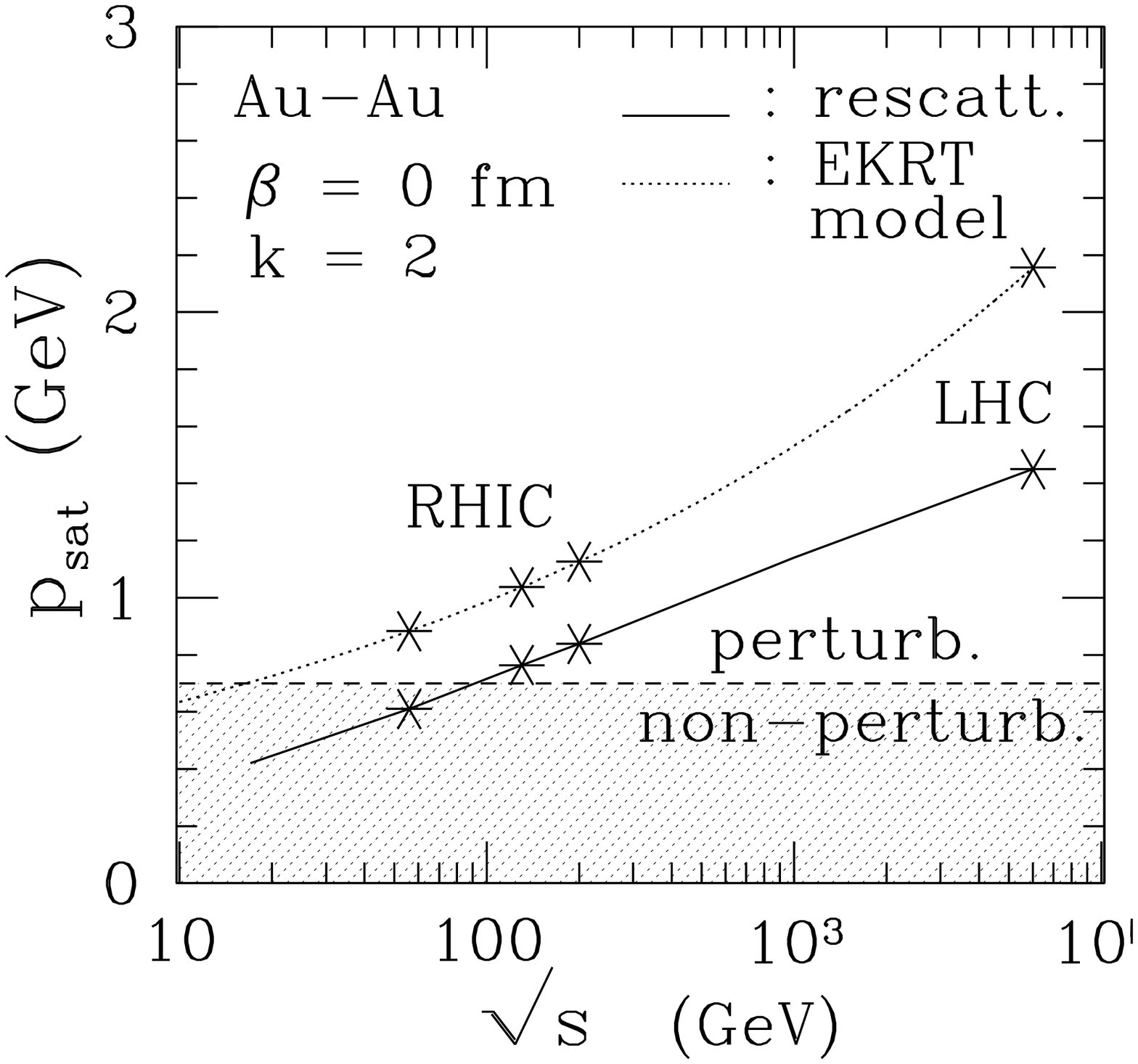,height=4.5cm}
c) \epsfig{figure=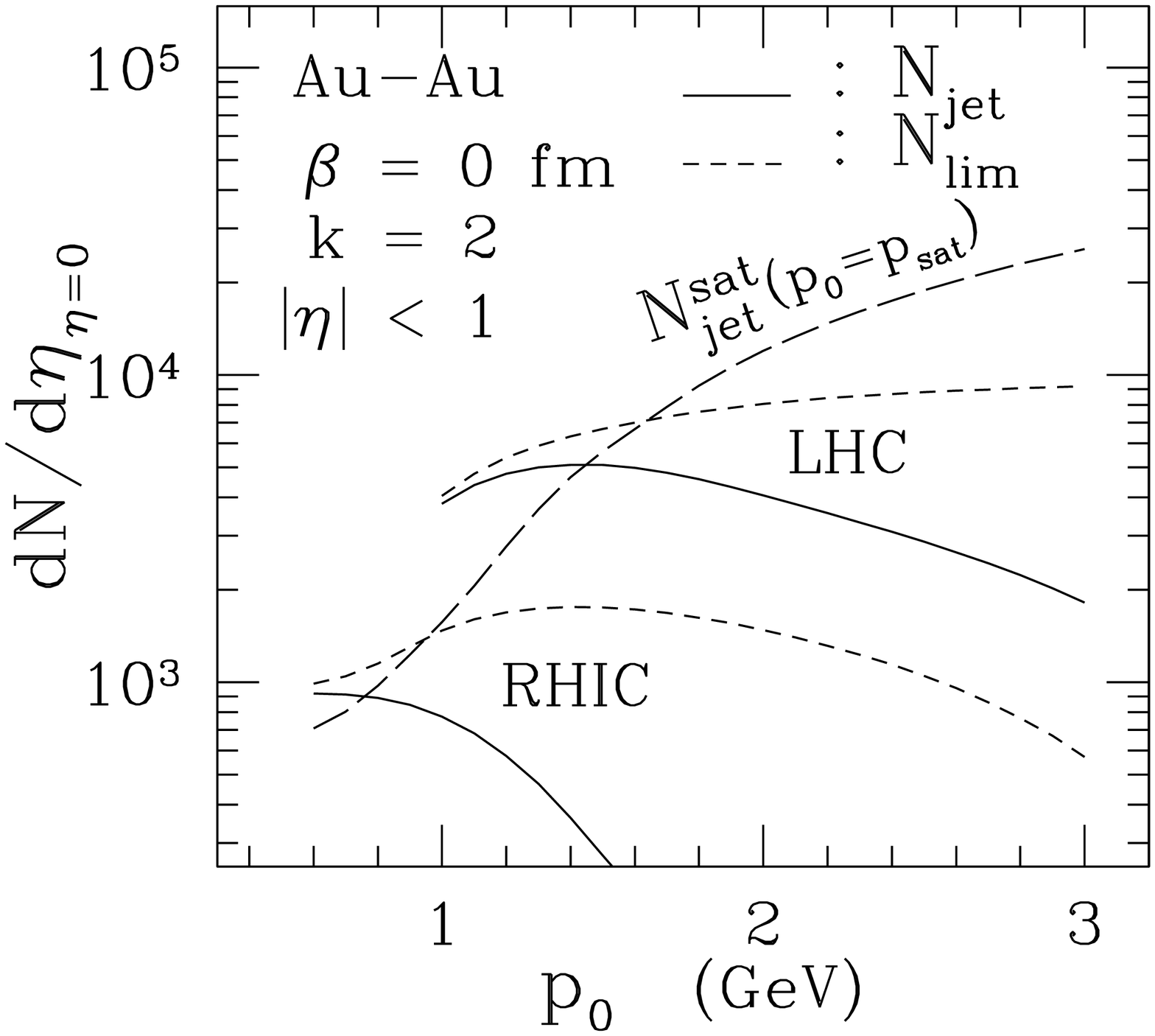,height=4.5cm}
\vskip -.2cm
\parbox{15.2cm}{\caption{a) Minijet multiplicity dependence on the
cutoff. Solid lines include rescatterings, dashed lines are the Eikonal
estimate. b)~Saturation cutoff dependence on the c.m. energy compared to the
EKRT model. c)~Saturation happens when the solid line crosses the long-dashed
one, see Eq.(\ref{psat}).} 
\label{fig:rescatterings} }
\end{center}
\vspace*{-.3cm}
\end{figure}

The simplest rescattering event is a three-parton elastic
scattering. It can be shown \cite{T} that in the limit $t/s \ra 0$
and for infinite number of colours this cross-section factorizes in
terms of two two-body elastic cross-sections. Therefore the process
can be pictured as two successive semi-hard scatterings against two
different target partons (see Fig. \ref{fig:collisions}b). We can try
to extend this picture to an arbitrary number of scatterings in the
following way \cite{CT}. We consider the {\em multi-parton distributions}
$D^A_n(\vec{x}, \vec{{b}})$, i.e. the probability densities of having $n$
partons in the A nucleus with fractional momenta $x_i$ and transverse
coordinates ${b_i}$. If we neglect correlations inside the nuclei the
multi-parton distributions are Poissonian:    
$	D^A_n (\vec{x}, \vec{b}) = {1\over n!}    
		\Gamma_A(x_1, {b}_1) \ldots   
		\Gamma_A(x_n, {b}_n)    
		\exp [-\int dx d^2b \, \Gamma_A(x, {b}) ]   $.   
Then, we consider the probability that at least one semi-hard interaction 
occurs between $n$ partons from A and $m$ partons from B:
$ P_{nm} = 1 - \prod_{i=1}^n \prod_{j=1}^m (1-\hat{\sigma}_{ij})$,
where $ \hat\sigma_{ij} = \sigma_H(x_ix_j)\delta^{(2)}(b_i-b_j)$ is the
probability of a semi-hard scattering between the $i$-th parton from A and the
$j$-th from B. One can then obtain the minijet 
cross-section by convoluting the multi-parton distributions and the semi-hard 
probability: 
$ \sigma_{mj} = \sum_{n,m=1}^{\infty} \int D_A^n P_{nm} D_B^m $. In this way 
we are describing both the disconnected collisions and the rescatterings 
(see Fig. \ref{fig:collisions}b). The minijet multiplicity is no longer 
related in a simple way to the number of collisions, however if we
choose a fixed scale $Q=p_{cut}$ for the distribution functions and
the running coupling constant, we can obtain a simple expression for it
\cite{CT}: 
\beq
	N_{jet}(\beta) = \int dxd^2b \, \Gamma_A(x,b-\beta) 
		\left(1-e^{-\int dx'\sigma_H(xx')\Gamma_B(x',b)} \right) 
		+ \{ A\leftrightarrow B\}_{\ds \stackrel{\lora}
		{p_{cut}\ra \infty}} N_{jet}^{eik}(\beta)
  \label{Njet}
\eeq
We interpret it as the density of projectile partons $\Gamma_A$ times the 
probability of having at least one semi-hard scattering, and the exponent 
in (\ref{Njet}) as the opacity of the target. As $p_{cut}$ goes to
infinity the target opacity becomes smaller and smaller, therefore the
projectile is traversing a very dilute target, the chance of scattering
twice or more becomes negligible and the minijet multiplicity
approaches the Eikonal estimate. On the contrary, as the cutoff
decreases the opacity increases, the target becomes almost black and  
the multiplicity reaches a finite limit: roughly speaking
\beq
	{N_{jets}}_{\ds \stackrel{\lora} 
		{\scriptstyle p_{cut}/\sqrt{s} \ra 0}} 
		N_{lim} =\int\Gamma_A \ .
  \label{Nlim}
\eeq
This basically expresses the fact that a collision cannot free more
partons than were present in the incoming nuclear wave-function. The
effect of the rescatterings is shown in Fig.\ref{fig:rescatterings}a,
where the minijet multiplicity tends to saturate below 2 GeV at LHC
and 1 GeV at RHIC. Note that the dependence on the cutoff has become much
smoother than in the Eikonal estimate, making the choice of the
cutoff less critical. An analogous behaviour is observed also in the
initial transverse energy and for the $k$-factor dependence
of these quantities \cite{AT}. 

It is possible to show that in (\ref{Njet}) we are taking into account
not only the jets that had $n \geq 1$ semi-hard scatterings, but also
those that had {\it at least} one semi-hard scattering but also any
number of soft ones \cite{T}. Indeed one semi-hard scattering is enough to
shadow completely the soft ones, so that in (\ref{Njet}) the soft cross-section
does not appear. In this way, we are neglecting only the minijets that
had just soft scatterings. This, together with (\ref{Nlim}), allows to
define a {\it saturation cutoff} $p_{sat}$: we require that the
number of jets be near the limit and we call $\psat$ the cutoff at
which this happens; for example we can require
\beq
	N_{jet}(p_{cut}=p_{sat})= 80\% \, N_{lim}(p_{cut}=p_{sat}) \ .
  \label{psat}
\eeq 
Finally, we compute the initial conditions at this cutoff: 
$\Njet^{sat}=\Njet(\pcut=\psat)$.

Now, we have to check if $\psat$ is far enough form $\Lambda_{QCD}$,
in which case we can trust these perturbative computations and say
that the semi-hard minijets saturate the initial production
mechanism. At RHIC energies this method is at the border of
applicability, and at LHC it should be reliably
applicable (Fig.\ref{fig:rescatterings}b). The percentage chosen in
(\ref{psat}) is an arbitrary parameter, however it happens that
$\psat$ sits where the plateau in the minijet multiplicity begins
(Fig.\ref{fig:rescatterings}c), so that choosing a different value
does not change too much the results.    

\begin{figure} 
\begin{center}
\epsfig{figure=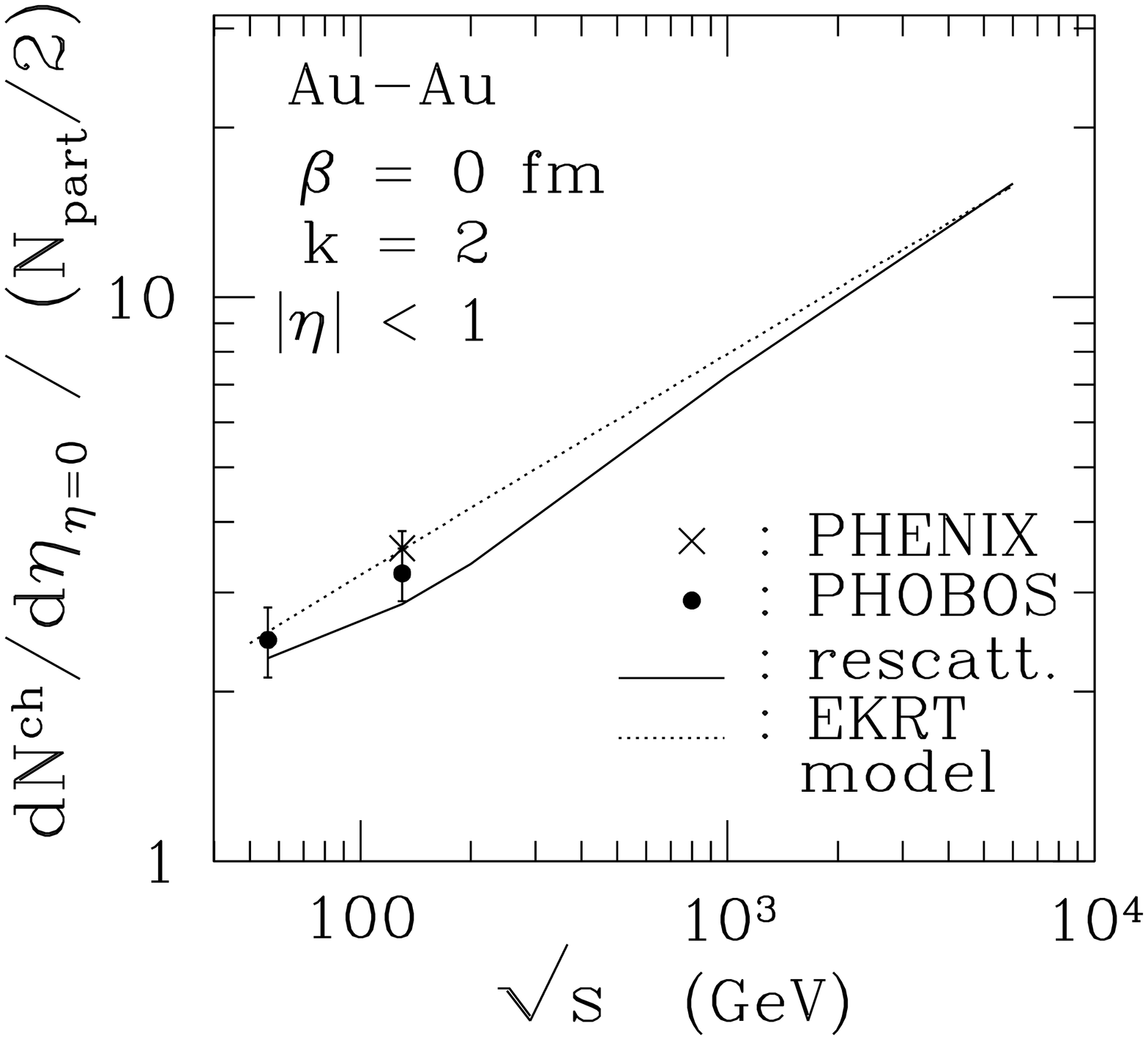,height=5.31cm}
\hspace*{1cm}
\epsfig{figure=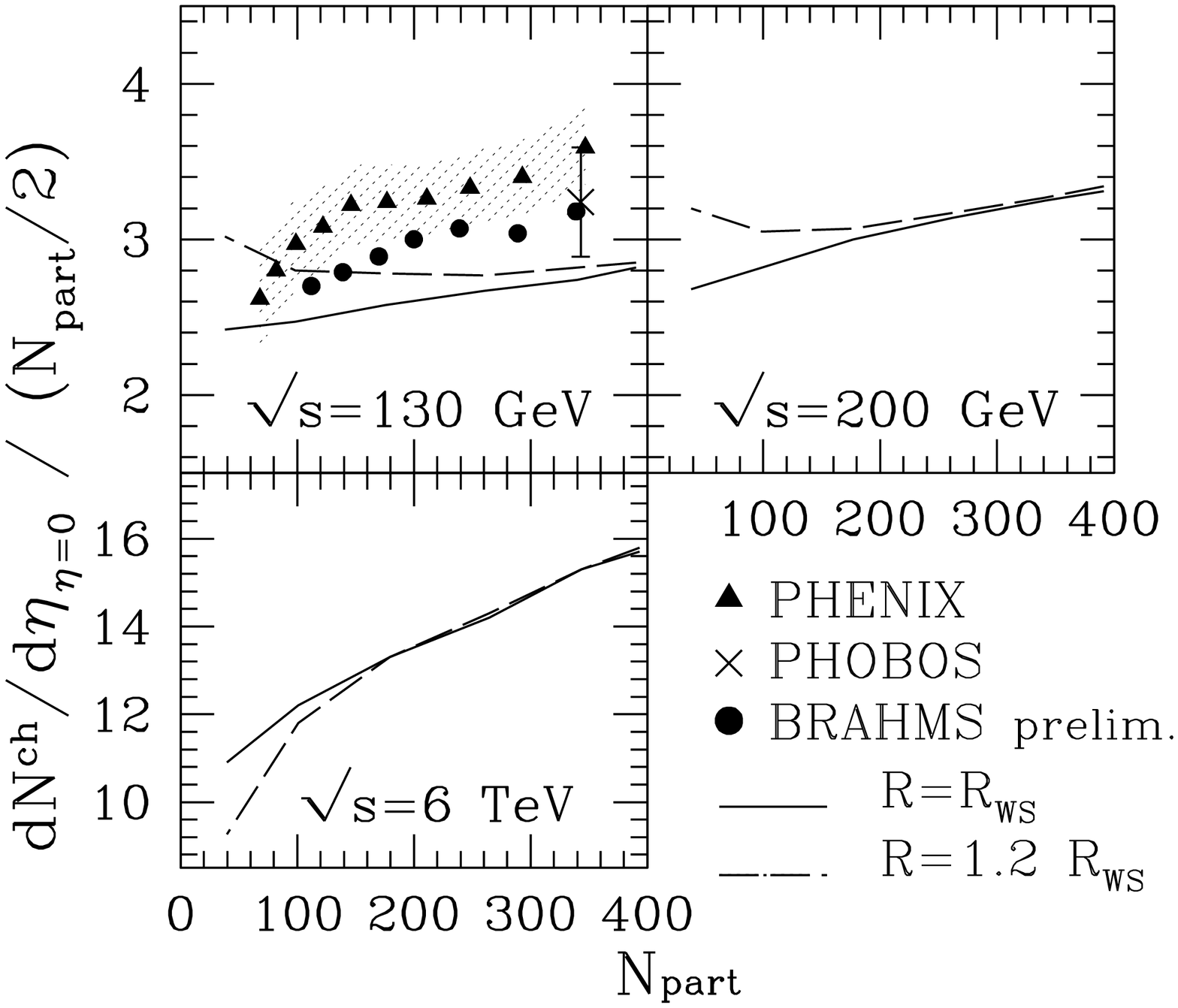,height=5.5cm}
\vskip -.2cm
\parbox{12cm}{\caption{Charged particle density at mid-rapidity per
participant nucleon pair. {\it Left:}~Center of mass energy dependence.
{\it Right:}~Centrality dependence. }
\label{fig:charged} }
\end{center}
\vspace*{-.2cm}
\end{figure}

With these tools we can study the charged particle multiplicity. We assume
isentropic expansion of the minijet plasma in the central rapidity region 
and direct proportionality between minijets and final hadrons \cite{EKRT}, 
$	N^{ch}(\sqrt{s},\beta)=0.9 \ {2 \over 3}
	\  N_{jet}^{sat}(\sqrt{s},\beta) \ , $
and compute the number of participant nucleons at fixed energy and centrality
in a Glauber model:
$	N_{part}  = \int  d^2b \, \tau_A(b-\beta)  
		\left(  1  - \exp[-\sigma_{pp}(\sqrt{s})\tau_B(b)]  \right) $, 
where $\sigma_{pp}$ is the inelastic proton-proton cross-section.
To study non-central collisions with Wood-Saxon thickness functions, we have
to enforce the collision geometry by assigning the nuclei a radius $R$
and by computing 
$\psat$ inside the overlap area defined  by this radius. The reason is that in
its periphery a nucleus is very dilute, therefore in that region it makes no
sense to impose that it becomes black. The
results are shown in Fig. \ref{fig:charged}a as a function of the
c.m. energy and of the number of participants. 

For central collisions the dependence on the energy is strikingly
similar to the EKRT saturation model \cite{EKRT}, however in our case
the saturation is determined dynamically and is related to the
blackness of the target, moreover no shadowing corrections have been
included in the parton distribution functions. For $R$
equal to the Woods-Saxon radius, $R_{WS}$, the slope of the
centrality dependence at $\sqrt{s}=130$ GeV compares very well with 
the experimental data \cite{exp} down to 180 participants; it 
underestimates {\sc phenix} data but is not bad compared to {\sc
brahms} and {\sc phobos}. Remember also that we are neglecting {\it ab
initio} up to 20\% of soft partons in the initial conditions
computation, and that the saturation is not expected to work well for
peripheral collisions. 
At this energy the dependence on the radius $R$ is rather
strong, but it decreases already at $\sqrt{s}=200$ GeV, and almost
disappears at LHC energies. Note also that the slope of the curve
increases with the energy signaling an increasing role played by
purely semi-hard events. This is in sharp contrast with the EKRT
model, where an almost flat behaviour is found.

In summary, the inclusion of the rescatterings in the dynamics of the
collision allows to define global observables also at very low values
of the cutoff, in a region of interface between perturbative and
non-perturbative interactions. The minijet multiplicity tends to saturate
and to reach a limit given by the number of partons in the initial
nuclear wave functions. This allows to define a dynamical saturation cutoff at
which the initial conditions can be estimated. The results compare
satisfactorily to RHIC data at $\sqrt{s}=130$ GeV, where the model is
at the edge of its applicability. As the energy increases it should
be more and more reliable. \vspace{-.7cm}\\

\section*{Acknowledgments}   
I wish to express my gratitude to D. Treleani and  U. Wiedemann 
for long and enlightening discussions. This work was
partially supported by the Italian ministry of university and of
scientific and technological research ({\sc murst})  under the grant 
{\sc cofin99}. \vspace{-.7cm}\\

\end{document}